\documentclass[a4paper,aps,prl,reprint,superscriptaddress, twocolumn,preprintnumbers,amsmath,amssymb,nobalancelastpage]{revtex4-1}

\usepackage{amsmath}
\usepackage{verbatim}
\usepackage{graphicx}
\usepackage{color}
\usepackage[normalem]{ulem}
\usepackage{amsmath,amsfonts,amssymb,graphicx,color,times,bbm,psfrag,dsfont,slashed,bm}
\usepackage{gensymb}
\usepackage{natbib}

\usepackage[T1]{fontenc}
\usepackage[latin9]{inputenc}

\usepackage{lipsum}
\usepackage[dvipsnames]{xcolor}

\begin{document}

\preprint{}

\title{Localization-Driven Quantum Sensing}

\author{Ayan Sahoo}
\affiliation{Harish-Chandra Research Institute, A CI of Homi Bhabha National Institute,
Chhatnag Road, Jhunsi, Allahabad 211 019, India}

\author{Utkarsh Mishra}

\thanks{Present address: Ming Hsieh Department of Electrical and Computing Engineering, University of Southern California, Los Angeles,  CA 90089, USA.}

\affiliation{Institute of Fundamental and Frontier Sciences,
University of Electronic Science and Technology of China, Chengdu 610054, China }

\author{Debraj Rakshit}
\affiliation{Harish-Chandra Research Institute, A CI of Homi Bhabha National Institute,
Chhatnag Road, Jhunsi, Allahabad 211 019, India}

\date{\today}

\begin{abstract}

We show that the delocalization-localization transition in  quantum-many body (QMB) systems are a compelling quantum resource for achieving quantum-enhanced sensitivity in parameter estimation.  We exploit the vulnerability of a near-transition QMB state against the parameter shift for devising efficient sensing tools. In this realm, the main focus of this work is to identify, propose and analyze experimentally relevant quantum observables for precision measurement. Taking a QMB system as a Fermi lattice under quasi-periodic modulation that supports an energy-independent delocalization-localization transition, we suggest operator-based adiabatic and dynamical quantum sensors endowed with considerable quantum advantages. In particular, we analyze single-particle system and the system at half-filling. While the quantum Fisher information saturates Heisenberg limit, we demonstrate that that experimentally relevant suitable observables with a spotlight on the charge-density-wave operator. Demonstrating their efficacy, these observables emerge as promising candidates for experimentally exploiting quantum advantages, showcasing superior performance compared to the standard quantum limit in terms of system-size scaling. Taking our exploration further into the dynamical realm, we discuss observables that yield super-extensive generic scaling in measurement precision. The comprehensive nature of our study not only sheds light on the intricacies of the delocalization-localization transition but also offers practical insights for the development of quantum sensors and their potential applications.
\end{abstract}

\maketitle


\emph{Introduction and motivation}.-- There are ongoing efforts to propose and engineer robust physical systems for quantum sensing. Successful experimental efforts addressing quantum enhanced
measurement precision in parameter sensing are atomic clocks~\cite{Appel09, Chauvet10}, interferometry~\cite{Mitchell04, Nagata07,LIGO13}, magnetometry ~\cite{Wasilewski10, Sewell12} and ultracold spectroscopy~\cite{Liebfried04, Roos06}. The ultimate limit of precision in parameter estimation is given by the quantum Cram\'er-Rao bound~\cite{Braunstein72,Cramer46,Helstrom76,degen2017}. The bound relates uncertainty, ${\cal E}_q$, in the estimation of an unknown parameter with the quantum Fisher information (QFI), $F_Q$ , as ${\cal E}_q{\geq}(MF_{Q})^{-1}$, where $M$ is the number of repetition of the sensing protocol.  QFI can reach the Heisenberg limit (HL), i.e.,  $F_Q{\sim}L^{2}$, for certain non-classical quantum states with $L$ qubits,  whereas it scales linearly with system size, $F_Q{\sim}L$, known as standard quantum limit (SQL), for $L$ independent qubits ~\cite{Giovann04,Giovann06}.

Quantum metrology covers a wide gamut of research topics starting from fundamental studies to applications in quantum technology
\cite{Pezze18review,Huelga97,Oszmaniec16,Zhou20,Naikoo23,Bhattacharyya22}. Recently, a new class of quantum sensing devices, termed adiabatic sensors, have surfaced. They exploit cooperative quantum phenomena in isolated quantum many-body systems, such as quantum phase transition~\cite{Sachdev,Dutta,Rams18}, for achieving Heisenberg scaling ~\cite{Zanardi08,Zanardi,Venuti07,Tsang13,Rams18,Garbe20,Mishra21,Gietka21,Chu21,Salado21,Lepori22,Montenegro21,Candia21,Mirkhalaf20,Pezze16,Pezze18,Roscilde18,You07,Zanardi07,Gu10,Braun18,Plenio22PRX,Plenio22Arxiv}. The motivation of exploring criticality for sensing stems from the fact that around the critical point the macroscopic quantum state changes drastically even with a small change in the parameter characterizing the external signal. It is worth to mention here that such approach of sensing has been employed for classical systems~\cite{classical_sensing}.
Other kinds of quantum phase transition beyond the Landau theory of spontaneous symmetry breaking can also be advantageous, e.g., topological transition~\cite{Bernevig13}, for achieving quantum enhanced sensing~\cite{ Sarkar22}.  In this realm, the localization-delocalization transition also harbors an enormous potential for quantum-enhanced parameter sensing, which this letter explores. The self-dual symmetry of the Aubry-Andr{\'e}-Harper (AAH)~\cite{Aubry80} model with quasi-periodic modulation leads to a localization transition as a finite modulation strength, $V_c$. Further works have confirmed the existence of a many-body localization transition in the presence of weak interactions~\cite{Michal14,Iyer13}. There the quantum fidelity susceptibility at the transition point scales as $L^{2/(d \nu)}$~\cite{Thakurathi12,Cookmeyer20}, where $d$ denotes  spatial dimension and the scaling exponent $\nu$ is associated with the localization length that  scales as $\zeta{\sim}|V - V_c|^{-\nu}$  with $\nu=1$~\cite{Sinha19,Wei19,Bu22}. It immediately implies that the sensitivity of the unknown  parameter $V$ scales as $L^{2/(d \nu)}$ at the transition point. However, as it is challenging to experimentally access the fidelity between neighbouring quantum states in QMB systems~\cite{Gu14,Zhang08,Zhang09}, the fidelity-based theoretical understanding does not promise an experimental translation~\cite{Cerezo2020}. Moreover, although a suitably optimized quantum operator, evaluation of which generally turns out to be non-trivial within the many-body setting,  can, in principle, saturate the maximum allowed precision governed by the QFI, it often transpires to be experimentally irrelevant~\cite{Zanardi08}. Instead, we need to focus on experimentally accessible observables that may overcome the shot-noise limit (SNL) by a sizable amount and provide a substantial quantum advantage. We wish to mention here a very recent work on Stark Localization~\cite{Wannier60}, which belongs to a different universality class, has been proposed for quantum-enhanced sensing, where the work is primarily focused on single-particle aspects of QFI~\cite{He23}.

\begin{figure*}
    \includegraphics[width=0.95\textwidth]{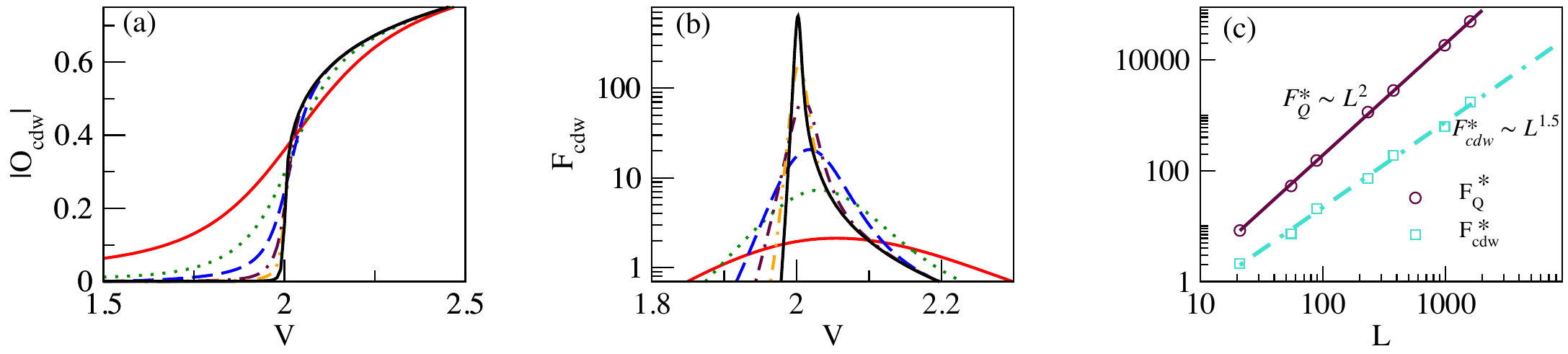}
\caption{{\bf Single particle case:} {\bf (a)} Modulus of $|O_{cdw}|$ as a function of $V$ for $L$ = 21 (solid-red), 55 (dotted), 89 (dashed), 233 (dashed-dotted), 377 (dashed-dotted-dotted), 987 (solid-black). The inflation points ($V^*$) corresponding to different $L$ approach the transition point, $V_c$, in the thermodynamic limit. {\bf (b)} OFI ($F_{cdw}$) corresponding to the operator $\hat{O}_{cdw}$ as a function of $V$ for the same set of system sizes presented in (a). The localization transition point at $V_c = 2$ is marked by the gradual prominence of the OFI peaks with increasing lattice sites. The plot styles are kept consistent.  {\bf (c)} The circles and the squares show the maximum QFI, $F_Q^{*}$, and maximum OFI, $F_{cdw}^{*}$ at $V^*$ as a function of the $L$. The straight line show the best fits for $F_Q^{*}$ (solid line) and $F_{cwd}^{*}$ (dashed line). QFI admits Heisenberg scaling, i.e., $F_Q^{*}{\sim}L^2$. The best fit is shown by the dashed straight line that reveals $F_{cdw}^{*}{\sim} L^{1.5}$.}
    \label{fig:fig1}
\end{figure*}
Moreover, we resort to dynamical sensing protocols via sudden quench to encode the unknown parameter $V$ in the probe state. Here the protocol duration time, $t$, manifests itself as a fundamental resource for the QFI scaling, along with the system size, $L$. Similar to the adiabatic considerations, sensing protocols have been designed via non-equilibrium dynamics~\cite{Polkovnikov11,Eisert15,Montenegro2022,Montenegro2023} influenced by criticality corresponding to a second-order quantum phase transition~\cite{Mishra21,Tsang13,Macieszczak16,Chu21,Mishra22,Rams18,Torrontegui,Garbe22}. In particular, under certain assumptions, the dynamical precision quantified via QFI can at-most scale as $L^2 t^2$, where $t$ represents total interrogation time~\cite{Boixo,Garbe20,Plenio22PRX}. This is the so-called HL in the dynamical context~\cite{Giovann04,Giovann06}. The dynamical sensors have certain advantages over the adiabatic counterparts, e.g., offering the protocol time as an additional resource for the scaling, overcoming the obstacles of critical slowing down, or designing better sensing protocols via the sudden quench strategy in comparison to the finite-time ramp required for implementing adiabatic protocols in reality~\cite{Garbe22, Yang22}. Moreover, the measurement precision associated with an observable may attain a super-extensive growth with respect to the interrogation time, which is beyond the quadratic scaling of QFI, while perfectly respecting the allowed Cram{\'e}r-Rao bounds.

\emph{Parameter estimation}.-- Here we provide essential background for estimating a single unknown parameter, $V$. We consider that the parameter is encoded either in the ground state or time-evolved state of a QMB system. As the parameter $V$ changes locally, its fluctuation is captured by the fidelity susceptibility, $\chi_Q$, defined by
\begin{equation}
 \chi_Q = -\lim_{\delta V \to 0}\frac{\partial^2 {\cal F}_Q}{\partial (\delta V)^2},
 \label{eq:fidelitysucc}
\end{equation}
where ${\cal F}_{Q} =  \langle \psi(V)|\psi(V+\delta V)\rangle$ is the quantum fidelity between two near-by quantum states $|\psi(V)\rangle$ and $|\psi(V+\delta V)\rangle$. 
The fidelity susceptibility is related to the QFI as $F_Q=4\chi_Q$~\cite{Zanardi08}. In general, one can infer the unknown parameter from the state $|\psi(V)\rangle$ by measuring an observable, $\hat{O}$. In the asymptotic limit, one can associate 
observable Fisher information (OFI),  $F_O$,  which is quantified by the error propagation formula followed from the inverse of signal-to-noise ratio (SNR), as \cite{Rams18,Pezze19}
\begin{equation}
\label{eq:error}
    F_O^{-1} =\lim_{\delta V \to 0} \frac{\langle {\hat{O}}^2 \rangle -\langle {\hat{O}} \rangle^2 }{\Big(\frac{d\langle {\hat{O}}\rangle}{d(\delta V)}\Big)^2}.
\end{equation}
The quantum Cram{\'e}r-Rao bound~\cite{Braunstein72} provides the bound on the uncertainty for any observable estimation: $F_O(V, \hat{O}) \le F_Q(V)$ \cite{Cramer46,Helstrom76,Pezze19}. In the rest of the present letter, we exploit the delocalization-localization transition, identify experimentally accessible observables with quantum advantage, and propose schemes for designing adiabatic and dynamical quantum sensors.

\begin{figure}
    \centering
    \includegraphics[width=8cm,height=13cm]{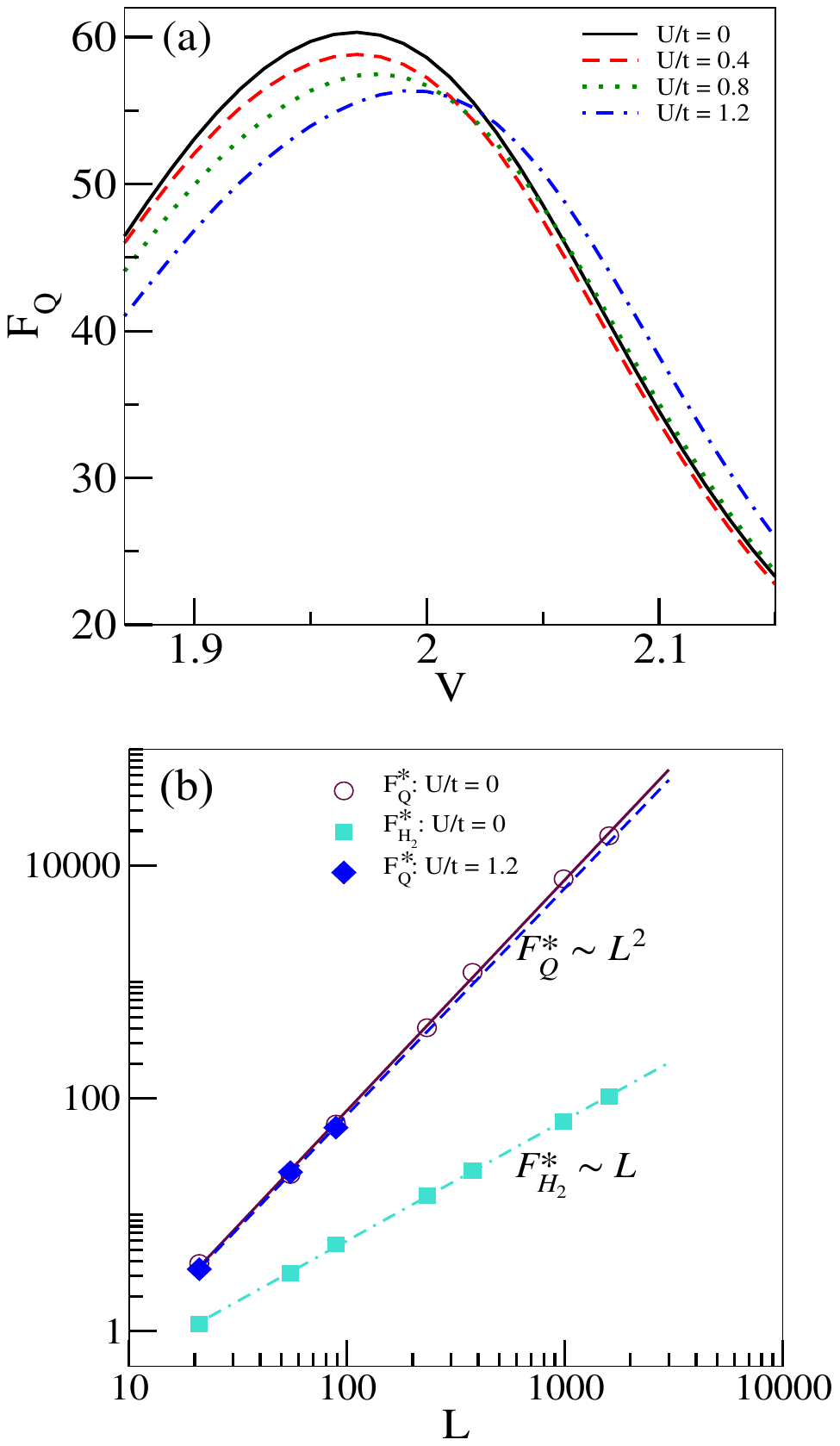}
    \caption{{\bf Half-filled case -- Adiabatic:} {\bf (a)} demonstrates interaction effects for particular case at half-filling ($L=89$ and $n_f=45$). The plot shows $F_Q$ as a function of $V$ for varied interaction strengths (up to the interaction strength comparable to the kinetic energy). {\bf (b)} The maximum QFI, $F_Q^{*}$, and OFI, $F_{H_2}^*$, corresponding to a natural operator, $\hat{H}_2$,  as a function of $L$ for $U{=}0$. The number of fermions $n_f$ for $L$ = 21, 55, 89, 233, 377, 987 and 1597 are 11, 28, 45, 116, 189, 494, and 798, respectively. While $F_Q^{*}$ nearly saturates the Heisenberg limit, i.e., $F_Q^{*}{\sim}L^2$ for both, the bare system and a system with moderate interaction ($U{=}1.2$, represented via diamonds),  $F_{H_2}^{*}$ scales linearly with $L$, and hence only saturates the SQL.}
    \label{fig:fig1}
\end{figure}
\emph{System}.-- We study a one-dimensional fermionic lattice with an underlying quasi-periodically modulated onsite potential. The generic form of the Hamiltonian in the context of single-parameter estimation is
 $\hat{H}{=}\hat{H}_1 + V \hat{H}_2$. For the system under consideration, $H_2$ represents a quasi-periodically modulated onsite potential. Let's consider the single-particle limit first, where $H_1$ has contribution only from the kinetic energy, $\hat{H}_1 \equiv \hat{H}_{ke}$. In this case, the final Hamiltonian is given by 
\begin{eqnarray}
\label{eq:Ham}
 \hat{H} = - \sum_{i}(\hat{c}^{\dagger}_{i}\hat{c}_{i+1}+h.c)+V \sum_{i} \cos(2\pi i \omega) \hat{c}^{\dagger}_{i}\hat{c}_{i},
\end{eqnarray}
where  $\omega$ is an irrational number, and $\hat{c}^{\dagger}_{i} (\hat{c}_i)$ is fermionic creation (annihilation) operator at the $i^{th}$ site. The Hamiltonian represents the AAH model, that has been studied in various contexts, such as the Hofstadter butterfly \cite{Hofstadter76}, transport \cite{Purkayastha18,Sutradhar19}, mobility edge \cite{Saha16,Ganeshan15,Ganeshan1502}, criticality \cite{Szabo18,Wei19,Rakshit21}, and topological phases\cite{DeGottardi13,Cai13,Fraxanet3}. The AAH model has a self-dual symmetry between the Hamiltonians in  the momentum and the position space, which leads to an energy independent localization transition at a finite modulation strength, $V_c=2$. For a given $V$, all the states are either localized (for $V > V_c$) or extended (for $V < V_c$).  However, we restrict our interest to the localization transition in the ground state, and examine its usefulness for adiabatic sensing. Due to the fractal nature and even-odd dichotomy, the system is not typical because all system sizes cannot fit under the same scaling functions \cite{Cestari11,Wei19}. For finite size systems, proper scaling emerges at the transition for system sizes $F_n$ with either odd or even sequences from the Fibonacci series and for $\omega$ to be approximated by $\omega_n{=}F_n/F_{n+1}$. Here $F_n$ and $F_{n+1}$ are two consecutive Fibonacci numbers with the property $\omega{=}\lim_{n\to \infty}(F_n/F_{n+1}) \to (\sqrt{5}-1)/2$, which is the so-called golden ratio. This work considers periodic boundary conditions (PBC), odd lattice sizes, and scaling at the transition point. 
Experimental observations of localization transition in the AAH model have been reported, e.g., in ultracold atom set-up \cite{Schreiber15,Negro03,Roati08,Modugno10,uab} and in photonic crystals \cite{Lahini09,Kraus12,Verbin13,Verbin15}.

In addition to the single-particle case, we study the half-filled case in the presence of interaction, $\hat{H}_1$ gets modified as $\hat{H}_1{=}\hat{H}_{ke}+\hat{H}_{int}$, where we choose to keep the interaction term simple nearest-neighbor type, such that
\begin{equation}
\hat{H}_{int} = U \sum_i \hat{n}_i \hat{n}_{i+1},
\end{equation}
where $\hat{n}_i=\hat{c}^{\dagger}_{i}\hat{c}_{i}$ and $U$ is the interaction strength.

The localization transition persists in the ground state in the presence of interaction \cite{Mastropietro15,Mastropietro17}. Moreover, further theoretical works have argued in favour of many-body localization (MBL) \cite{Iyer13,Khemani17,Naldesi16,Setiawan,Michal14}, albeit with different universality properties than the uncorrelated disordered system.

\emph{Single Particle case: Adiabatic}.-- 
The QFI, defined via Eq.~(1), saturates the HL, i.e., it scales quadratically with the system size: $F_Q^* (V=V^*) \sim L^2$, where $V^*$ corresponds to quasi-periodic potential amplitude for which QFI maximizes.   
It reveals the potential of the AAH-type many-body Hamiltonian to be exploited for building a new class of quantum sensing devices. We now need to identify experimentally accessible observables endowed with quantum advantage.

We focus on an observable that measures the occupation imbalance between the even and odd sites. The corresponding operator is ${\hat{O}_{cdw}}{=}\sum_{i}(-1)^{i} \hat{c}^{\dagger}_{i} \hat{c}_{i}/n_f$, which reveals the charge density wave (CDW) order in a quantum state. For instance, this operator is directly measurable within the optical lattice set-up loaded with ultracold atoms \cite{Schreiber15}. We consider system size up to $L=1597$. The computation of OFI, as followed from Eq.~(2), requires evaluation of the expectation values of $\langle \hat{O}_{cdw} \rangle$, $\langle \hat{O}_{cdw}^2 \rangle$, where $\hat{O}_{cdw}^2{=}\sum_{i,j} (-1)^{i+j} \hat{n}_i \hat{n}_j$.  From now on, we switch to a more convenient notation for the operator expectation value: $O_{\alpha} \equiv \langle \hat{O}_{\alpha} \rangle$.  Figure~1(a) shows the modulus of the charge density order, $|O_{cdw}|$, as a function of the amplitude of the potential, $V$. Due to the extended nature of the quantum state,  $O_{cdw}$ gradually diminishes with increasing $L$ in the delocalized phase but acquires a finite value in the localized phase. This feature, as expected, becomes sharper with increasing system size and proceeds to assume a step function-like structure with vanishing $|O_{cdw}|$ in the extended phase in the thermodynamic limit. The inflation points corresponding to different system sizes tend to converge towards the infinite system transition point, $V_c{=}2$. Figure 1(b) shows OFI, $F_{cdw}$, corresponding to the various system sizes as a function of $V$. The $F_{cdw}$ develops a peak at finite size transition points, $V^*$. They are characterized via gradual prominence of the peaks at near $V_c$. While QFI at the transition point, $F_Q^*$, admits a scaling of $F_Q^*{=}L^{2.01(1)}$, the OFI at the transition point, $F_{cdw}^*$,  scales like  $F_{cdw}^*{=}L^{1.54(4)}$ (see Fig.~1(c)). The scaling associated with $F_{cdw}^*$ is less than the HL limit but comfortably beats the SNL, hence offering a genuine quantum advantage in the measurement precision.

\emph {Half-filled case: Adiabatic.}-- We now turn our attention to the 
the half-filled case, i.e., the number of fermions is $(L\pm1)/2$ \cite{suppli1}, with PBC. We perform density matrix renormalization group (DMRG) calculations via matrix product state (MPS) formalism for the interacting systems  \cite{White04,Schollwock11}. 
We probe the ground state $F_Q$ for system sizes upto $L{=}1597$ $(n_f {=}798)$ for the non-interacting (NI) cases and upto $L{=}89$ $(n_f{=}45)$ in the interacting cases~\cite{pbc}. The influence of interaction remains minimal on the scaling properties for weak to moderate interaction strengths \cite{pbc}.

Figure 2(a) presents $F_Q$ as a function of $V$ for different interaction strength $U$ and fixed $L$. We note two observations: First, the peak tends to shift towards a higher $V^*$ with increasing $U$ -- This is expected as many-body localization transition is supposed to occur at $V_c > 2$ in the presence of interaction; and second, the value of $F_Q$ tends to slightly decrease at $V^*$ with increasing $U$. The scaling of the QFI at $V^*$, $F_Q^*$ is presented in Fig.~2(b). $F_Q^*$ scales as $F_Q^*(U{=}0){=} L^{1.98(2)}$ in the NI limit, i.e., it saturates the HL limit. Despite the lack of enough data points, it is evident that the effects of interaction on the scaling exponent remain pretty small in the range of weak to intermediate range. For the case shown here, QFI also nearly saturates the HL limit: $F_Q^* (U=1.2){=}L^{1.95(4)}$. 

Proposing an experimentally relevant observable with quantum advantage for fractionally filled equilibrium case turns out to be non-trivial. The operator $\hat{O}_{cdw}$ is not a suitable operator as the ground state is devoid of CDW (charge-hole) ordering in the localized phase \cite{suppli1}. Unlike the single particle case, it fails to characterize the transition, and no proper scaling with $L$ is found. However, a proper scaling can be found with other suitable observables, such as $\hat{O}_{H_2}{=}\sum_{i} \cos(2\pi i \omega) \hat{c}^{\dagger}_{i}\hat{c}_{i}$ (which is basically $\hat{H}_2$, but a new notation has been introduced for consistency: $\hat{O}_{H_2}  \equiv \hat{H}_2$). The corresponding OFI ($F_{H_2}$), as followed from Eq.~(\ref{eq:error}) admits a proper scaling with $L$: $F_{H_2}^* = L^{1.04(1)}$, but it barely beats the SQL.  One may think of a few strategies, one of which is by preparing the system artificially in a highly excited state with high CDW order and then monitoring the adiabatic evolution of $\hat{O}_{cdw}$ by tuning $V$ to lower values. This strategy, however, may not be useful because of closely lying near-degenerate excited states or energy crossings. The second way is to adopt a dynamical strategy for engineering dynamical sensors. The protocol is to prepare the system in the ground or highly excited state with high CDW order and then quench to or through $V_c$ to either of the phases. 

\begin{figure}
    \centering
    \includegraphics[width=0.45\textwidth]{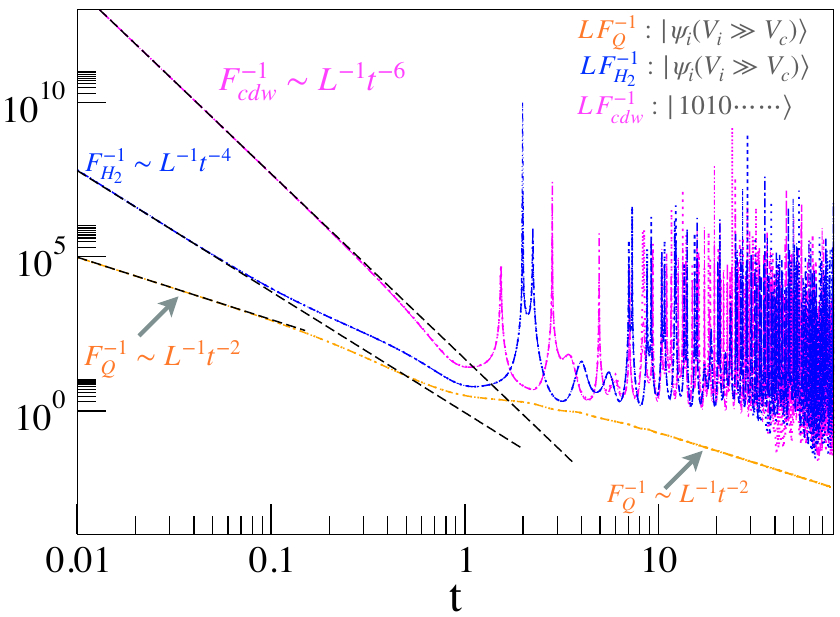}
    \caption{{\bf Half-filled case -- Dynamics:} Time dependence of QFI and OFI due to a sudden quench in $V$ ($U{=}0$) for $L${=}55 (dot-dot), 89 (dash-dot) and 233 (dash-dot-dot). Collapse plots of the inverse of the quantities are presented after rescaling them with $L^{\alpha}$. Considering an initial state as the ground state prepared in the localized phase (case shown is for $V=5$), the figure shows scaled QFI (orange lines).  The scaled data set collapses for $\alpha{=}1$, implying  $F_Q{\sim}L t^2$. OFI ($F_{H_2}$), presented for the same initial state, scales as $F_{H_2}{\sim}L t^4$ (blue lines) at short or transient time. Finally, the OFI, $F_{cdw}$, corresponding to an initial state with maximum CDW order is shown to scale like, $F_{cdw}{\sim}L t^6$ (magenta lines) for $t{\lesssim}1$. Data corresponding to different $L$ merge on top of each other, and the naked eye may be unable to differentiate them.}
    \label{fig:fig1}
\end{figure}

\emph {Half-filled case: Dynamics.}--  Two specific schemes are considered:  Performing a sudden quench from an initial state, $|\psi_{in}\rangle$, which can be i) the ground state corresponding to a particular phase, or ii) a state with maximum CDW order ($O_{cdw}{=} 1$), e.g., $|1010\cdots\rangle$, which is a high excited state of the system in the limit $V \to \infty$. Such a high CWD state can be prepared artificially with high fidelity, e.g., in ultracold atom experiments \cite{Schreiber15}. Further scheme to construct QFI runs as follows: i) Pump a continuous resource for a specific duration, $t$, via unitary dynamics provided by a driving Hamiltonian, $\hat{H}(V_f)$, ii) back propagate the evolved state, $| \psi(V_f,t) \rangle$, with the Hamiltonian $\hat{H}_f(V_f+\delta V)$, iii) the back propagated state is projected on the initial state in order to obtain the fidelity, ${\cal F}(t){=}\langle \psi(V,t) |\psi(V+\delta V,t) \rangle$. Importantly, as there may be an experimental limitation on time due to decoherence, the observables' short- or transient-time behavior is of prime interest.

We illustrate the time dependence of the QFIs and OFIs calculated under the above mentioned situations in Fig.~3 for $U {=} 0$ and different system sizes $L{=}$ 55, 89, and 233 at half-filling. We plot the inverse of the quantities under consideration and rescale them with $L$. Data sets corresponding to different $L$'s collapse quite well, implying a saturation in the SNL as a function of $L$. In Fig.~3, we show $F_Q^{-1}$ and $F_{H_2}^{-1}$ for the initial state, $|\psi_{in}(V>>V_c)\rangle$, as the ground state prepared  in the localized phase and then driven to the extended phase. While $F_Q$ overall scales as $\sim Lt^2$ in the short and long times \cite{QICproof,Brun,Duer,QGua}, two characteristic timescales emerge in  OFI's. At short/transient time, roughly upto $t{\sim}1$, $F_{H_2}$ assumes a higher scaling in $t$: $F_{H_2}{=}L t^{3.92(1)}$. The observables exhibit rapid oscillations at $t{\gtrsim}1$. We find the long time average value to have a roughly quadratic scaling in time. However, $\hat{O}_{cdw}$ holds prime interest, as described in the context of adiabatic sensors. Considering the charge-hole ordered state $|1010\cdots\rangle$ as the initial state, the dynamics allow us to circumvent the scaling issues associated with the adiabatic scenario. It even offers superior scaling in the transient time in comparison to $F_{H_2}$: $F_{cdw}{=}L t^{6.003(7)}$. Note that although the $F_O$ exhibits a higher scaling than $F_Q$ in transient time, the absolute value of $F_O$ is upper bounded by $F_Q$, as expected. Remarkably, when the system is not prepared too close to the transition point,  the dynamical scaling laws are quite generic, i.e., they are independent of the final driving Hamiltonian and the initial state \cite{suppli1}.

\emph{Discussions}.-- In this work, we perform canonical equilibrium and dynamical studies of single-parameter quantum sensing via experimentally accessible natural observables in a quasi-periodically modulated single-particle and half-filled Fermi lattice. Important findings are the followings: Within the adiabatic scenario, while the QFI can attain HL limit in both, the single-particle and half-filled cases, an experimentally accessible observable, CDW order ($\hat{O}_{cdw}$), is proposed that is endowed with significant quantum advantage at localization transition in the single particle limit. The same observable, however, is not suitable in the half-filled case for the scaling analysis, as typical ground state configuration needs an important ingredient, the charge-hole ordering. Instead, the system can be initially prepared in a quantum state that maintains the required ordering and then can be quenched across or at the transition point. Remarkably, this provides a superextensive \emph {generic} (i.e., robust against the final driving Hamiltonian) scaling in the measurement precision. We also report another natural observable, $\hat{O}_{H_2}$, experimental access for which may be comparatively non-trivial as it requires measurement of fermionic occupation number at individual sites, that provides significant quantum advantages in precision measurement even if the system is initially prepared in the ground state. In conclusion, we propose localization-delocalization transition as a resource that can be exploited for engineering a new class of quantum sensors. This idea brings the scope for exercising further exciting investigations, e.g., in the context of multi-parameter sensing, sensing via partial accessibility, or sensing via localization induced by uncorrelated disorder.

\acknowledgements
D.R. acknowledges support from Science and Engineering Research Board (SERB), Department of Science and Technology (DST), under the sanction No. SRG/2021/002316-G.

\onecolumngrid

\begin{thebibliography}{99}
\bibitem{uab} AAH model is particularly suitable for implementing via ultracold atoms in optical lattice by effectively freezing the dynamics the two-orthogonal directions by employing deep lattice in those directions and introducing an additional detuning lattice that is much weaker in comparison to the primary lattice
\bibitem{pbc} As we impose periodic boundary condition (PBC), the convergence in the MPS bond dimension is more costly compared to that of open boundary condition (OBC). The truncation error achievable by $m$ bond dimension in OBC requires $m^2$ bond dimension in PBC. The interacting analysis is naturally restricted to small system sizes, and few available data points available from the Fibonacci series are not sufficient for a proper scaling analysis. However, through a comparative analysis between the NI and the interacting systems in context of QFI, we find that the role of interaction remain minimal, while fermionic statistics, the manifestation of which comes in form of a proper anti-symmetrization of the single-particle eigenstates, plays the major role in the observed many-body effects.
\bibitem{QICproof} It can be shown that the time-evolution of QFI under a generic Hamiltonian of the form, $\hat{H}{=}\hat{H}_1 + V \hat{H}_2$, has a quadratic dependence in time: $F_Q{\sim}t^2\Big(\langle \overline{(\mathcal{\hat{H}}_2)}^{2}_{t}\rangle- |\langle\overline{(\mathcal{\hat{H}}_2)}_t\rangle|^2\Big)$, where $\overline{(\mathcal{\hat{H}}_2)}_t{=}\frac{1}{t}\int_{0}^{t}\hat{H_{2}}(t')dt'$~\cite{Brun, Duer, QGua}.
\end{thebibliography}
\end{document}